\documentclass[letterpaper,showpacs,preprint,secnumarabic,preprintnumbers,amsmath,amssymb,floatfix,showkeys]{revtex4}
\usepackage{graphicx}% Include figure files
\usepackage{subfigure} %% fuer \subfigure
\usepackage{amssymb}
\usepackage{fancyhdr} %%fuer Running Title

\usepackage[format=default,justification=raggedright,labelformat=simple,font=footnotesize]{caption}
\DeclareCaptionLabelSeparator{dot}{.\quad }
\bibpunct{[}{]}{;}{a}{,}{,}
\newcommand{\shorttitle}{Delay-induced~multistability}

\begin{document}

%%Erzeugen der Kopfzeile mit dem running title
\pagestyle{fancy}
\fancyhead[r]{\shorttitle}
%%

%\preprint{submitted to Int. J. Bifurcation and Chaos (2007)}

\date{\today}

\title{Delay-induced multistability near a global bifurcation}

\author {J. Hizanidis}
\author {R. Aust}
\author {E. Sch\"oll}

\affiliation{Institut f{\"u}r Theoretische Physik, Technische Universit{\"a}t Berlin,  Hardenbergstra{\ss}e 36, D-10623 Berlin, Germany}

\begin{abstract}
We study the effect of a time-delayed feedback within a generic model for a saddle-node bifurcation on a limit cycle. Without delay the only attractor below this global bifurcation is a stable node. Delay renders the phase space infinite-dimensional and creates multistability of periodic orbits and the fixed point. Homoclinic bifurcations, period-doubling and saddle-node bifurcations of limit cycles are found in accordance with Shilnikov's theorems.
\end{abstract}

	%nonlinear dynamical systems
\maketitle

\section{Introduction}
Time-delayed feedback was originally proposed in the context of chaos control \citep{PYR92} as an alternative to the famous OGY method developed earlier by \citet{OTT90}. The idea was to achieve stabilization of unstable periodic orbits (UPOs) by adding, to a chaotic system, a control force in the form of the difference between a system variable at time $t$ and at a delayed time $t-\tau$.
 This method proved to be very powerful and has been successfully applied to various physical systems since 
then \citep{SCH07}. The scheme was improved by \citet{SOC94}, and other variants have 
been elaborated \citep{KIT95,BAB02,BEC02,UNK03,SCH03a}, and applied also to 
stochastic systems \citep{GOL03,JAN03,HAU06}. 
Moreover, elegant analytical theories \citep{JUS97} were developed 
supporting, thus, numerical findings.  Apart from the practically relevant application 
of time-delayed feedback, e.g. to lasers \citep{SCH06a}, 
the interest lies highly on the mathematical aspect of the 
problem. Delay differential equations are difficult to handle. 
The delay renders the system infinite-dimensional and the interplay with nonlinearity 
uncovers complex dynamic behaviour.
Delay-induced multistability was already predicted in the first paper by \citet{PYR92}. 
The idea that time-delayed feedback may not only be used for controlling a system but 
also for creating new dynamics is not new. Nevertheless, the investigation of 
delay-induced bifurcations and multistability is still a growing field \citep{XU04,BAL05} with applications to the logistic map as well as to laser equations \citep{MAR03,MAR05a}.

\section{The model}
Here we study the influence of time-delayed feedback in a generic model representative 
for excitable dynamics \citep{GAN93,DIT94,RAP94}.
The behaviour of the system without delay is governed by a global bifurcation, namely a 
saddle-node bifurcation on a limit cycle ({\em saddle-node infinite period bifurcation,
SNIPER}), that takes place when a certain parameter exceeds a threshold. Such a
bifurcation was first observed experimentally in a semiconductor device \citep{PEI89},
and also encountered in various semiconductor models, e.g. for Gunn domains 
\citep{SHI96}, superlattices \citep{PAT98,HIZ06}, or lasers \citep{DUB99,WIE02,KRA03a}. 
Excitability, naturally, yields the system highly sensitive to fluctuations and therefore this system 
has served perfectly as an example for {\it coherence resonance} \citep{PIK97} shown 
in the seminal paper by \citet{GAN93} and \citep{DIT94} over ten years ago. In this 
paper we do not consider the effect of noise but extend the generic model by 
incorporating time-delayed feedback according to the Pyragas scheme. 
The equations are the following: 

\begin{eqnarray}
  \label{eq:model}
\dot{x}&=&x(1-x^2-y^2)+y(x-b)-K(x-x_{\tau})\\
\dot{y}&=&y(1-x^2-y^2)-x(x-b)-K(y-y_{\tau}).
\end{eqnarray}
Here $x$ and $y$ denote the variables at time $t$, while $x_{\tau}$ and $y_{\tau}$ the 
delayed ones at time $t-\tau$, with $\tau$ and $K$ 
being the delay and control strength, respectively. This kind of control is called {\it 
diagonal} because the control force may be written in the form of a unity matrix. 
In the absence of delay, i.\,e.\ $K=0$, $b$ plays the role of the bifurcation paramater. 
In polar coordinates $x=r \cos{\varphi}, y=r \sin{\varphi}$ Eq.~(\ref{eq:model}) with $K=0$ reads

\begin{eqnarray}
  \label{eq:polarcoord}
\dot{r}&=&r\left( 1-r^2 \right)\\
\dot{\varphi}&=&b-r\cos\varphi
\end{eqnarray}
When $b<1$ there are three fixed points: an unstable focus at the origin and a 
pair of a saddle-point and a stable node on the unit circle with coordinates $(b,+\sqrt{1-b^2})$ and $(b,-\sqrt{1-b^2})$, respectively. 
The latter collide at $b=1$ and a limit cycle $r=1$ is born. Above but close to the bifurcation, 
the frequency $f$ of this limit cycle obeys a characteristic square-root scaling law
$f \sim (b-1)^{1/2}$. 

\section{Linear stability analysis of the delay equation}
We prepare the system slightly below the bifurcation ($b=0.95$) and switch on the control. 
The first question that arises concerns the stability of the three fixed points and how 
this changes, or not, due to delay. For this, we perform a standard linear stability 
analysis and derive the characteristic equation for the roots, $\Lambda$, which 
determine the stability of the fixed points. 
For the unstable focus, the characteristic equations is:

\begin{equation}
  \label{eq:char_poly_ufo}
(1-K+Ke^{-\Lambda \tau}-\Lambda)^2+b^2=0.
\end{equation}

Due to the presence of the delay, Eq.~(\ref{eq:char_poly_ufo}) has infinitely many solutions.
However, the stability of the fixed points is determined by a finite number of critical 
roots with largest real parts.
Using the Lambert function $W$, which is defined as the inverse function of $g(z)=ze^z$ for 
complex $z$ \citep{HAL71,COR96}, the solution of Eq. (\ref{eq:char_poly_ufo}) can be 
expressed as:

\begin{equation}
  \label{eq:Lambert_ufo}
\Lambda=\frac{1}{\tau}W[K\tau e^{\tau(K-1\pm b)}]-K+1\mp ib.
\end{equation}

\begin{figure}[htbp]
  \centering
  \includegraphics[width=\columnwidth]{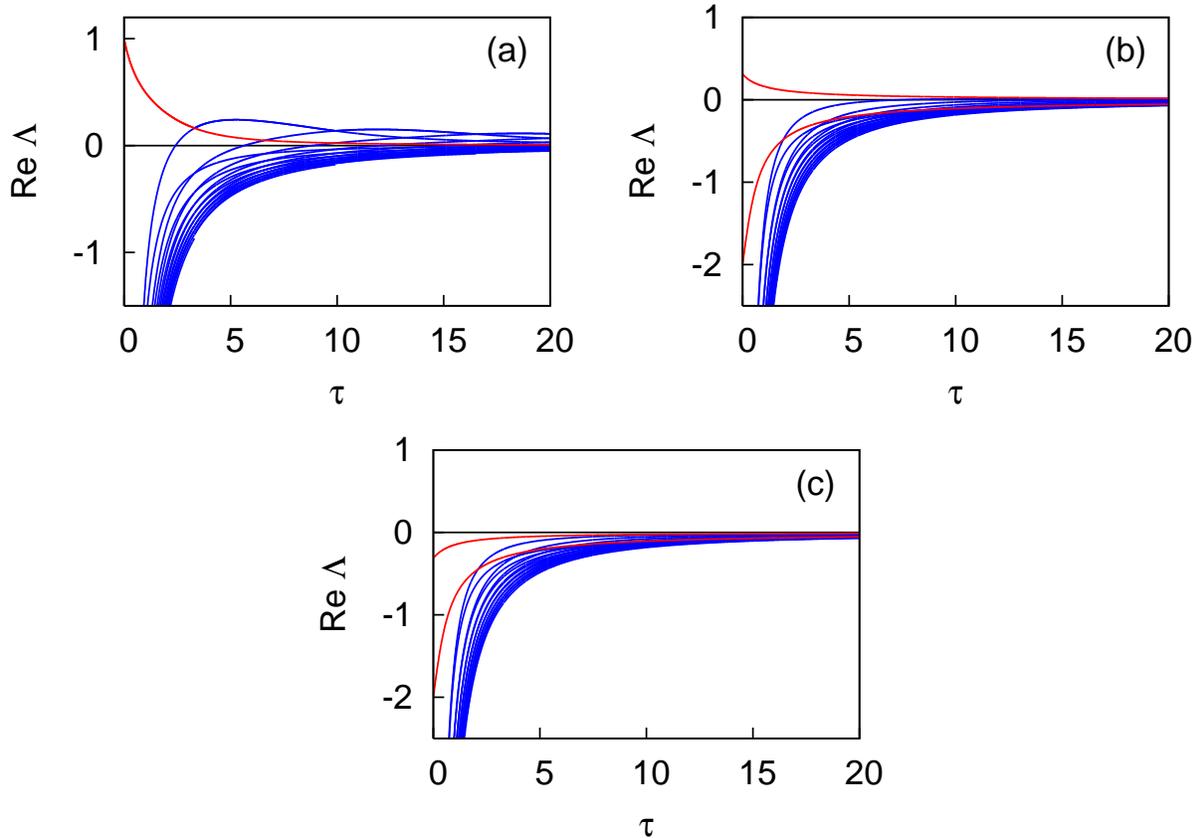}
\caption{Real parts of the complex eigenvalues $\Lambda$ as a function of $\tau$, for fixed $K=1$ and $b=0.95$. For the (a) unstable focus, (b) the saddle point and (c) the stable node. The modes emerging from the uncontrolled system and the delay-induced modes are marked red and blue, respectively.}
 \label{fig:real_eigen_vs_tau}
\end{figure}

In the case of the saddle and the node, the characteristic equation can be factorized into two equations:

\begin{eqnarray}
  \label{eq:char_saddle_node}
\Lambda+K+2-Ke^{-\Lambda \tau}&=&0\\
\Lambda^{s,n} +K \mp \sqrt{1-b^2} -Ke^{-\Lambda^{s,n} \tau}&=&0,
\end{eqnarray}

with solutions:

\begin{eqnarray}
  \label{eq:Lambert_saddle}
\Lambda_1&=&\frac{1}{\tau}W[K\tau e^{\tau(2+K)}]-2-K\\
\label{eq:Lambert_node}
{\Lambda_2}^{s,n}&=&\frac{1}{\tau}W[K\tau e^{\tau(K\mp \sqrt{1-b^2})}]-K\pm \sqrt{1-b^2}.
\end{eqnarray}
The superscripts '$s$' and '$n$' denote the saddle (upper sign) and the 
node (lower sign), respectively. 
Figure~\ref{fig:real_eigen_vs_tau} shows the real parts of the eigenvalues $\Lambda$ as a 
function of $\tau$ for a fixed value of $K$ for all three fixed points. One 
may see the eigenvalues of the uncontrolled system at $\tau=0$, and their interaction 
with the delay-induced modes (blue) with increasing $\tau$. In all three cases, control is 
unable to change the stability of the fixed point: in the case of the unstable focus, the 
mode with the largest real part (red) tends to very small values with increasing delay, 
remaining however positive. The same holds for the unstable mode of the saddle. 
Symmetric behaviour is observed for the stable modes of both saddle and node: 
they tend to zero as a function of $\tau$ but remain negative.
The picture does not change qualitatively even for other values of $K$ and therefore one 
might conclude that no delay-induced bifurcations of fixed points take place.  

\section{Global bifurcation analysis}
\begin{figure}[htbp]
  \centering
  \includegraphics[clip,width=\columnwidth]{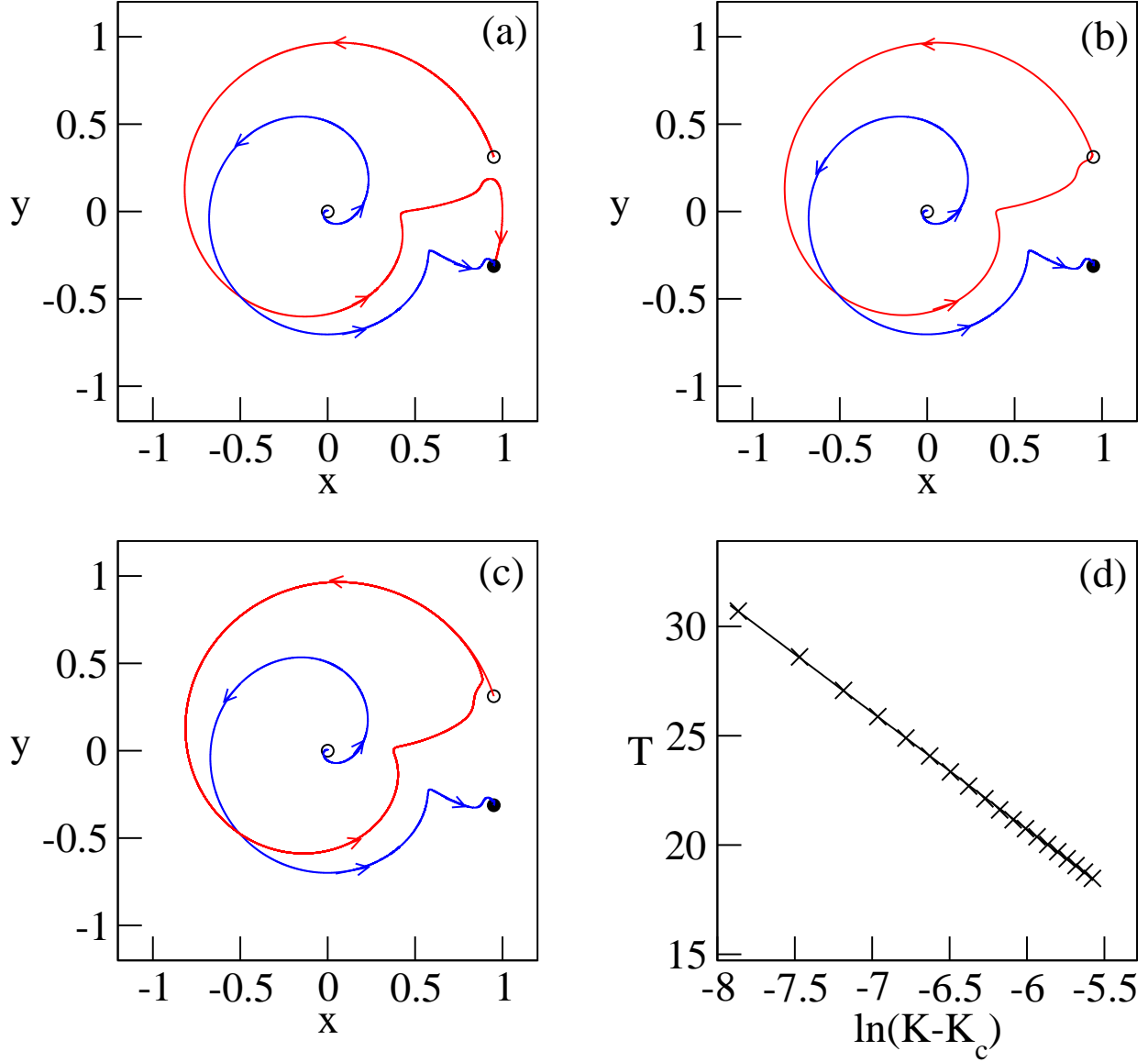}
 \caption{(a) Two dimensional projection of the phase space below the homoclinic 
bifurcation ($K=0.335$). (b) Homoclinic orbit (red) achieved at $K_c=0.3401$. 
(c) Delay-induced limit cycle (red) above the homoclinic bifurcation ($K=0.3438$). 
(d) Scaling of the oscillation period $T$ above but close to the critical point $K_c$
(crosses: simulation data, solid line: linear fit).
Full and open circles mark stable and unstable fixed points, respectively.
Parameters: $b=0.95$, $\tau=3$. }
  \label{fig:hom_scaling_law}
\end{figure}

However, the above local analysis gives no information on the global changes in phase space that delay potentially induces.
A numerical investigation shows, in fact, that there exists bistability in a certain parameter regime in the $K$-$\tau$ plane: trajectories starting  close 
to the saddle point are attracted by a delay-induced limit cycle, whereas trajectories starting elsewhere end up in the stable node.
Keeping $\tau=3$ fixed we find the critical value $K_c$ of $K$ for which this delay-induced 
limit cycle is born and observe a scaling $T \sim \ln\left\vert K-K_c\right\vert$ in the period $T$ of the 
corresponding oscillations, typical for the case of a homoclinic bifurcation. In 
Fig.~\ref{fig:hom_scaling_law} phase portraits of the system below, at and above the 
bifurcation are shown. Trajectories with different initial conditions are shown: 
one starting from the vicinity of the unstable focus in the origin (blue) and one from 
the vicinity of the saddle (red). Bistability is revealed in Fig.~\ref{fig:hom_scaling_law}(c) where two attractors 
(the stable node and the delay-induced limit cycle) coexist. The 'kink' in the trajectory 
shortly before the loop closes is due to the control: the control force starts acting at 
$t=3$ when the system is still moving on the slow part of the unit circle. 
Therefore, its effect is not so noticeable. As the system moves faster the 
control force attains higher values and the trajectory starts deviating from its deterministic path at $t=13$.
This deviation becomes large at $t=18$ where the trajectory, shortly before settling in the stable node, appears to be "attracted" to the saddle, resulting in this 'kink' in the $x-y$ projection.
%resulting thus in this delayed 'kink' in the phase 
%portrait. One could add that due to control, the trajectory deviates from its 
%deterministic path and in the $x-y$ projection appears to be ''attracted'' to the saddle. 
Also, as $K$ approaches the critical value, the trajectory passes closer and closer to 
the saddle on its way to the stable node.  This ends in a homoclinic orbit at $K=K_c$,  
(Fig.~\ref{fig:hom_scaling_law}(b)) from which a periodic orbit  is generated 
(Fig.~\ref{fig:hom_scaling_law}(c)). The period of the born limit cycle scales 
according to $T\approx -{\Lambda_u}^{-1}ln|K-K_c|$, where $\Lambda_u$ is the real part
 of the least unstable eigenvalue of the saddle point (i.\, e.\,, the one closest to the 
imaginary axis). One may calculate this from Eq. (\ref{eq:Lambert_saddle}) and find 
${\Lambda_u}^{-1}=0.1739^{-1}=5.75$. This is in rather good agreement with the slope of the solid line in Fig.~\ref{fig:hom_scaling_law}(d) which equals $5.35$.

\begin{figure}[htbp]
\centering
\includegraphics[width=\columnwidth]{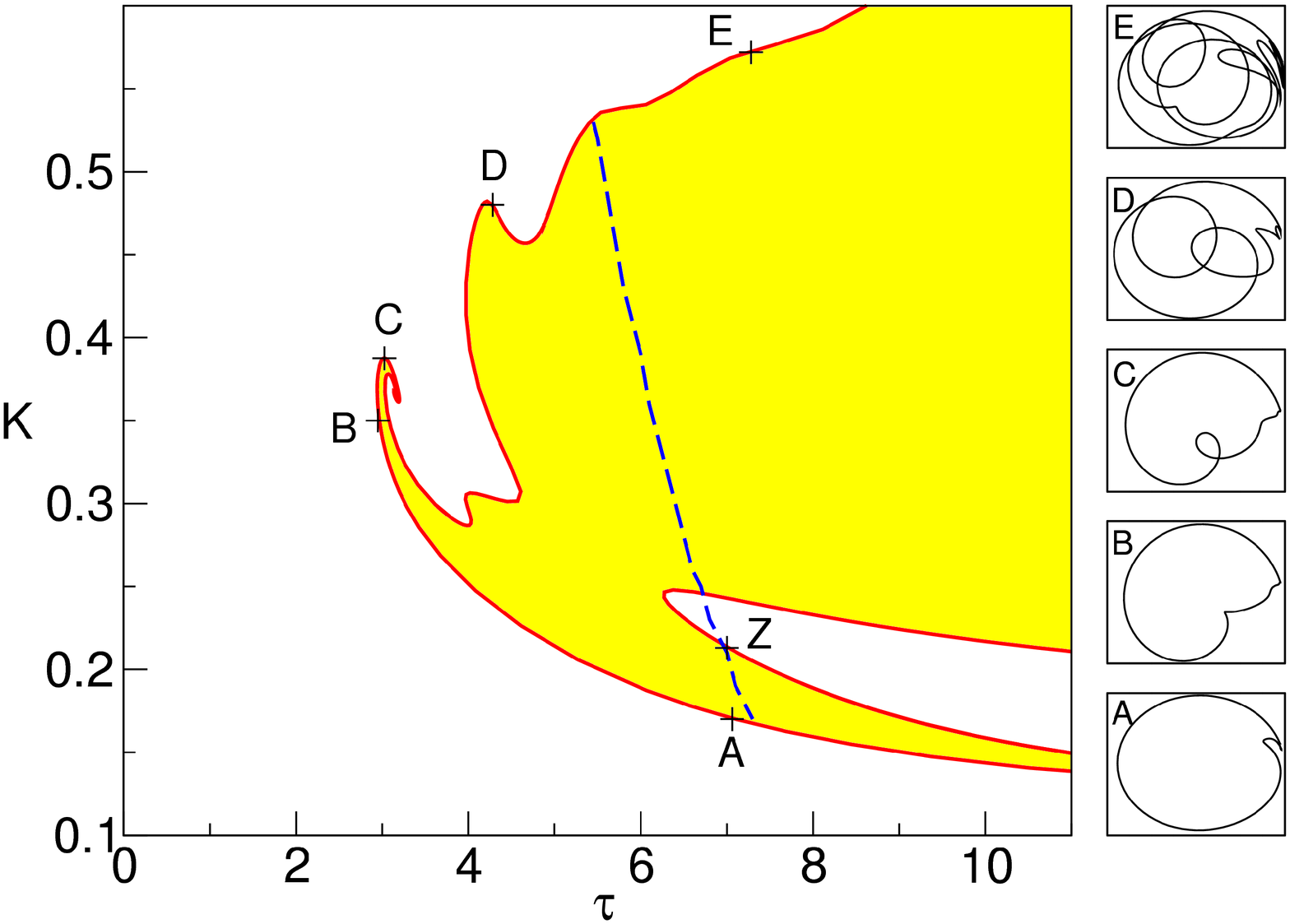}
\caption{Curve of homoclinic bifurcations (red) in the $K-\tau$ plane (left). 
A - E labels various points with homoclinic orbits, which are shown in the $x-y$ phase 
plane in the panel on the right. Delay-induced limit cycles exist, in addition to the
stable fixed point, in the yellow area. The blue dashed curve separates the regions $\sigma_0<0$ (left) and $\sigma_0>0$ (right)}
 \label{fig:hom_branch}
\end{figure}

In the following we use a bifurcation continuation tool 
\citep{ENG02,ENG01} and follow the homoclinic bifurcation in the $K-\tau$ plane. The 
produced bifurcation curve can be seen in Fig.~\ref{fig:hom_branch} (left). 
It consists of two main curves: one running through points $A-E$ and a second tongue-like curve.
In the white area the system is monostable (stable fixed point) while in the yellow area a delay-induced periodic
 attractor is born via a homoclinc bifurcation marked by the red curves.
%Crossing the curve from left to right the system becomes bistable as a delay-induced 
%periodic attractor is born via a homoclinc bifurcation. 

\begin{figure}[htbp]
  \centering
  \includegraphics[width=\columnwidth]{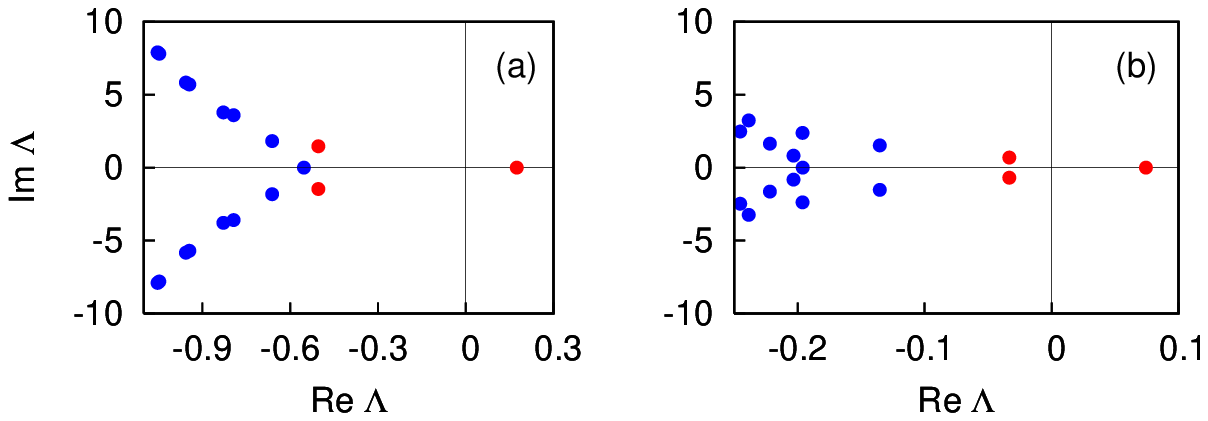}
 \caption{Spectra of the eigenvalues $\Lambda$ of the saddle for two points on the 
homoclinic bifurcation line: (a) $K=0.3401$, $\tau$=3 
(B in Fig.~\ref{fig:hom_branch}) and (b) {\bf $K=0.57$, $\tau=7.28$} (E in Fig.~\ref{fig:hom_branch}). 
The saddle quantity $\sigma_0=Re(\Lambda_s)+Re (\Lambda_u)$ is negative in (a)
whereas in (b) it is positive. ($b=0.95$)}
  \label{fig:spectra}
\end{figure}

At this point one should emphasize the role of the saddle-point: due to the delay, the 
saddle possesses no longer two distinct eigenvalues (one positive, i.\,e.\ unstable, and one 
negative, i.\,e.\  stable) but infinitely many. Moereover, complex eigenvalues come into 
play as well. The eigenvalues, however, that determine the behaviour of the colliding 
homoclinic orbit, are the leading ones, i.\,e.\  those closest to the imaginary axis. In 
Fig.~\ref{fig:spectra} one can see the eigenvalue spectrum for two parameter values 
on the homoclinic curve and notice that the leading eigenvalues of the saddle (red) are
a positive real eigenvalue, as in the original uncontrolled system, and a complex 
conjugate pair with negative real parts, generated by the delay.  This means that the 
saddle may turn into a {\it saddle-focus} for certain values of $K$ and $\tau$. 
Homoclinic orbits attached to a saddle-focus approach the fixed point in an oscillating 
manner. This explains the phase portraits in Fig. \ref{fig:hom_branch} (right) which become more and more complicated as $K$ increases.

\begin{figure}[tbp]
\centering
\includegraphics[clip,width=\columnwidth]{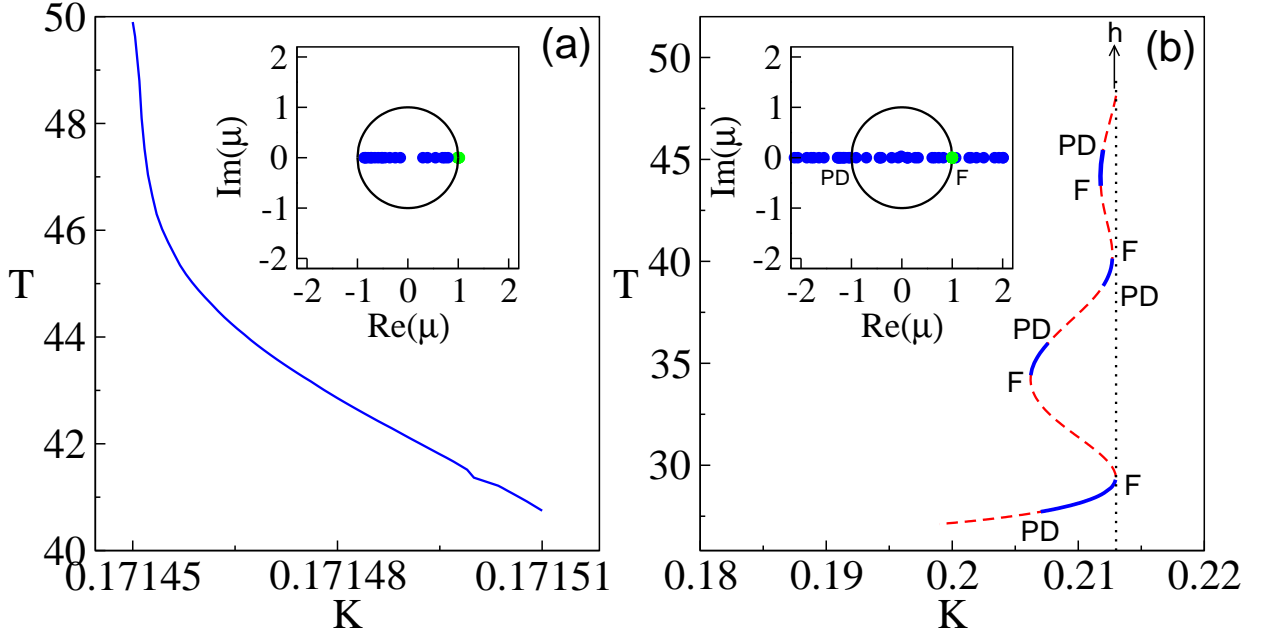}
\caption{ (a) Period $T$ of limit cycle born in a homoclinic bifurcation at $(K,\tau)=(0.17145,7)$ (point A in Fig.~\ref{fig:hom_branch}, $\sigma_0<0$). (b) Period $T$ of limit cycles in the multistable regime at $(K,\tau)=(0.213,7)$ (point Z in Fig.~\ref{fig:hom_branch}, $\sigma_0>0$), undergoing infinitely many fold (F) and period-doubling (PD) bifurcations, before ending in a homoclinic orbit h for $T\rightarrow\infty$ at $K=0.213$.
Solid blue and red dashed lines denote stable and unstable limit cycles, respectively. 
The insets show the two leading Floquet multipliers of the periodic orbit 
$\mu_1=1$ (green) and $\mu_2$ (blue) with $K$ as a parameter in (a), and $T$ as a parameter in (b), in the complex plane. $b=0.95$ 
}
 \label{fig:tau7}
\end{figure}

From the above, it is clear  that the two basic ingredients responsible for the 
delay-induced dynamics in our system are the homoclinic orbits and the saddle-foci. 
The theory of homoclinic bifurcations for ordinary differential equations has been well 
developed \citep{KUZ95,GUC86,WIG88}.
As already mentioned, global bifurcations are strongly related to excitability and 
therefore one expects to encounter them in excitable systems. 
Various physical systems such as modulation-doped semiconductor heterostructures 
\citep{DOE92}, semiconductor lasers \citep{WIE02,KRA03a}, 
neuron models \citep{FEU00} and chemical systems \citep{BOR06} have been studied in this 
respect, both theoretically and experimentally. On the other hand,  less work has been 
carried out for systems with delay undergoing such nonlocal bifurcations \citep{SET06}. 
It is therefore appropriate to analyze a generic system like the one studied here 
which, despite its simplicity, exhibits rich delay-induced dynamics with a homoclinic 
bifurcation as key component.

\section{Delay-induced multistability}
In what follows we will apply the theorems on homoclinic orbits connecting to saddle-foci as developed by Shilnikov \citep{KUZ95}.
According to them, the so-salled saddle quantity is crucial for the homoclinic 
bifurcations occurring in high-dimensional systems. The saddle quantity is defined as 
$\sigma_0=Re(\Lambda_s)+Re (\Lambda_u)$, where $\Lambda_s$ and $\Lambda_u$ are the leading 
stable and unstable eigenvalues, respectively. Shilnikov proved, among other, that 
negative $\sigma_0$ results in the birth of a unique stable limit cycle from a homoclinic 
orbit. On the other hand, for $\sigma_0>0$, a wide variety of homoclinic bifurcations 
may occur, some of which involve infinitely many periodic orbits in the vicinity of the 
homoclinic orbit. The green dashed curve in Fig.~\ref{fig:hom_branch} shows the condition $\sigma_0=0$. 
Along the homoclinic bifurcation line in Fig.~\ref{fig:hom_branch}
the saddle quantity changes sign, thereby allowing for both scenarios to take place. 
Figure~\ref{fig:spectra} shows the eigenvalue spectra for 
two different points on the bifurcation line corresponding to negative and positive 
saddle quantities, respectively. 

In the following, we restrict ourselves to a fixed value of $\tau=7$ and reveal 
multistability beyond the homoclinic bifurcation.
For $\tau=7$, a homoclinic bifurcation takes place at $K=0.17145$ (point A in 
Fig.~\ref{fig:hom_branch}). In this case $\sigma_0=-0.0116$, and the bifurcation creates 
one stable limit cycle, with Floquet multipliers within the unit circle (inset of 
Fig.~\ref{fig:tau7}(a) ). The period $T$ of this limit cycle increases monotonically as the 
bifurcation point is approached (Fig.~\ref{fig:tau7}(a)). 
Moving further along $\tau=7$ other homoclinic bifurcation curves are crossed, e.\,g.\ at $K=0.213$
(Fig.~\ref{fig:tau7}(b), cf. point Z in Fig.~\ref{fig:hom_branch}). 
There, the saddle quantity is positive ($\sigma_0=0.0023$, calculated analytically from 
Eq. (\ref{eq:Lambert_node})), and the picture is much more complicated: an infinite 
number of bifurcations take place, which are related to saddle-node (fold) bifurcations 
of pairs of stable and unstable limit cycles, and additional period-doubling (flip)
bifurcations of the stable limit cycles. The dependence of 
the period $T$ of the limit cycles upon $K$ is a nonmonotonic multivalued function,
whose turning points are associated with saddle-node bifurcations. In between the
fold bifurcations of the stable limit cycles, pairs of forward and inverse 
period doubling bifurcations occur.
In the insets of Fig.~\ref{fig:tau7} the trivial Floquet multiplier $\mu_1=1$ and the leading Floquet multiplier $\mu_2$ of the periodic orbit are plotted. It can be seen how $\mu_2$ changes as $T$ is varied along the multivalued function in the main figure, showing how fold and flip bifurcations occur, at $\mu_1=1$ and $\mu_1=-1$, respectively.
One should also expect other bifurcations near the critical point due to secondary 
homoclinic orbits which are beyond the scope of this paper.

\section{Conclusions}
In conclusion, we have presented a mechanism for delay-induced multistability in a 
system near a global bifurcation. In addition to the fixed point attractor which the 
uncontrolled system already possesses, a time-delayed feedback in the form of Pyragas 
difference control induces one or more coexisting 
limit cycle attractors. Depending upon
the feedback control strength $K$ and the delay time $\tau$, either a single stable
limit cycle is born in a homoclinic global bifurcation, or an infinite number of (stable and
unstable) periodic orbits is induced undergoing a rich menagerie of bifurcation scenarios
including period doubling and fold bifurcations. 
We have shown that the key ingredient in the observed dynamics is a homoclinic orbit 
connected to a saddle-focus created by delay. A bifurcation continuation in the 
$K-\tau$ plane was performed. Moreover, we were able to verify Shilnikov's theory of 
homoclinic bifurcations in a certain parameter regime. The excitable nature of the 
system and the infinite-dimensional phase space, due to delay, appear to play a 
crucial role in the induced homoclinicity. 

These results are interesting also from the point
of view of applications, since our generic model is representative 
for a wide range of real-world systems. For instance, the transition from stationary to 
moving field domains in semiconductor superlattices has been shown to be associated with 
a saddle-node bifurcation on a limit cycle as described by Eq.(\ref{eq:model}) at $K=0$
\citep{HIZ06}, and time-delayed feedback control can also be realized in this system 
\citep{SCH03a}. Already without delay, this system has been noted for its high 
multistability of stationary domain states \citep{PRE94,KAS94,KAS96}, and bistability
or higher multistability has been found in many other semiconductor nanostructures, 
see e.g. \citet{MEI00b,SCH01}.
    
\begin{acknowledgments}
This work was supported by DFG in the framework of Sfb 555. We are grateful to 
A. Balanov, G. Bordyugov, and G. Stegemann for fruitful discussions. 
\end{acknowledgments}

%\bibliographystyle{ijbc}
% \bibliography{hiz07_revision.bib}
%\bibliography{ref}

\end{document}